\documentclass[reprint,superscriptaddress,amsmath,amssymb,aps]{revtex4-1}
\usepackage{graphicx}% Include figure files
\usepackage{dcolumn}% Align table columns on decimal point
\usepackage{bm}% bold math
\usepackage{ulem} %includes strike out text \sout{}
\usepackage{siunitx} %for doing units properly
\usepackage{hyperref}% add hypertext capabilities
\hypersetup{colorlinks=true, linkcolor=blue, filecolor=magenta, urlcolor=cyan}
\usepackage[colorinlistoftodos]{todonotes} %add comments feature
\usepackage{color}

\begin{document}
%-----------------------------------------------------------------------%
% WORD COUNT ON OVERLEAF:
% By default, references, headers, captions, floats, displayed math etc are not included in the count. 
% I have excluded the abstract below using a tag, but NEED TO ADD CAPTIONS, EQUATIONS & FIGURES
% FROM PRL - limit is 3750 words  https://journals.aps.org/authors/length-guide
% Math; The word equivalent for displayed math is 16 words per row for single-column equations. Two-column equations count as 32 words per row.
% Figures; To estimate the word equivalent for figures use the figure’s aspect ratio (width / height). The estimate is [(150 / aspect ratio) + 20 words] for single-column figures, and [300 / (0.5 * aspect ratio)] + 40 words for double-column figures.
% The 4 figures we have are (conservatively);
% (1) Two-Col, 2792x636 = [(300 / (0.5*4.4) + 40] = 180
% (2) One-col, 1676x614 = [(150 / 2.73) + 20] = 80
% (3) One-col, 1986x1766 = [(150 / 1.12) + 20] = 160
% (4) One-col, 1708x1236 = [(150 / 1.38) + 20] = 130
% Captions are around (36+37+57+19)=137 in total.
% Therefore the total word contribution is +700
%-----------------------------------------------------------------------%

%-----------------------------------------------------------------------%
% Title & Authors
%-----------------------------------------------------------------------%
\title{Single-shot multi-keV X-ray absorption spectroscopy using an ultrashort laser wakefield accelerator source}
%setting up commands for each affiliation
\newcommand{\JAI}
{The John Adams Institute for Accelerator Science, Imperial College London, London, SW7 2AZ, UK}
\newcommand{\GOLP}
{GoLP/Instituto de Plasmas e Fus\~{a}o Nuclear, Instituto Superior T\'{e}cnico, U.L., Lisboa 1049-001, Portugal}
\newcommand{\CUOS}
{Center for Ultrafast Optical Science, University of Michigan, Ann Arbor, Michigan 48109-2099, USA}
\newcommand{\Lund}
{Department of Physics, Lund University, P.O. Box 118, S-22100, Lund, Sweden}
\newcommand{\Lancaster}{Physics Department, Lancaster University, Lancaster LA1 4YB, United Kingdom}
\newcommand{\CLF}{Central Laser Facility, STFC Rutherford Appleton Laboratory, Didcot OX11 0QX, UK}
\newcommand{\HZDR}{Helmholtz-Zentrum Dresden-Rossendorf, Bautzner Landstrasse 400, 01328 Dresden, Germany}
\newcommand{\ASCR}{Institute of Physics of the ASCR, Na Slovance 1999/2, 182 21 Prague, Czech Republic}
\newcommand{\TUD}{Technische Universit{\"a}t Dresden, D-01069 Dresden, Germany}
\newcommand{\LLNL}{Lawrence Livermore National Laboratory (LLNL), Livermore, California 94550, USA}

\author{B.~Kettle}
\email{b.kettle@imperial.ac.uk}
    %\homepage{http://www.Second.institution.edu/~Charlie.Author.}
    %\altaffiliation[Also at ]{Physics Department, XYZ University.}%Lines break automatically or can be forced with 
\affiliation{\JAI}

\author{E.~Gerstmayr}
\affiliation{\JAI}

\author{M.J.V.~Streeter}
\altaffiliation[Current address: ]{\JAI}
\affiliation{\Lancaster}

%alphabetical from here?
\author{F.~Albert}
\affiliation{\LLNL}    

\author{R.A.~Baggott}
\affiliation{\JAI}

\author{N.~Bourgeois}
\affiliation{\CLF}

\author{J.M.~Cole}
\affiliation{\JAI}

\author{S.~Dann} 
\altaffiliation[Current address: ]{\CLF}
\affiliation{\Lancaster}

\author{K.~Falk}
\affiliation{\HZDR}
\affiliation{\TUD}
\affiliation{\ASCR}

\author{I.~Gallardo Gonz\'{a}lez}
\affiliation{\Lund}

\author{A.E.~Hussein}
\affiliation{\CUOS}

\author{N.~Lemos}
\affiliation{\LLNL}

\author{N.C.~Lopes}
\affiliation{\GOLP}

\author{O.~Lundh}
\affiliation{\Lund}

\author{Y.~Ma}
\altaffiliation[Current address: ]{\CUOS}
\affiliation{\Lancaster}

\author{S.J.~Rose}
\affiliation{\JAI}

\author{C.~Spindloe}
\affiliation{\CLF}

\author{D.R.~Symes}
\affiliation{\CLF}

\author{M.~\v{S}m\'{i}d}
\affiliation{\HZDR}

\author{A.G.R.~Thomas}
\affiliation{\CUOS}
\affiliation{\Lancaster}

\author{R.~Watt}
\affiliation{\JAI}

\author{S.P.D.~Mangles} 
\affiliation{\JAI}
\email{stuart.mangles@imperial.ac.uk}

%-----------------------------------------------------------------------%
% Abstract
%-----------------------------------------------------------------------%
% The following tag ignores the section (until tag closed) for word counting
%%TC:ignore
\begin{abstract}

Single-shot absorption measurements have been performed using the multi-keV X-rays generated by a laser wakefield accelerator.
A 200 TW laser was used to drive a laser wakefield accelerator in a mode which produced broadband electron beams with a maximum energy above 1 GeV and a broad divergence of $\approx15~\textrm{miliradians}$ FWHM.
Betatron oscillations of these electrons generated $1.2\pm0.2\times10^6$ photons/eV in the 5 keV region, with a signal-to-noise ratio of approximately 300:1.
This was sufficient to allow high-resolution XANES measurements at the K-edge of a titanium sample in a single shot.
We demonstrate that this source is capable of single-shot, simultaneous measurements of both the electron and ion distributions in matter heated to eV temperatures by comparison with DFT simulations.
The unique combination of a high-flux, large bandwidth, few femtosecond duration X-ray pulse synchronised to a high-power laser will enable key advances in the study of ultra-fast energetic processes such as electron-ion equilibration.

\end{abstract}
%%TC:endignore
\maketitle

%-----------------------------------------------------------------------%
%   Introduction
%-----------------------------------------------------------------------%
% A Letter must begin with an introduction that states the issues it addresses and its primary achievements in language understandable across physics subfields

The extreme conditions present in high-energy-density (HED) matter make it notoriously difficult to study experimentally in the laboratory~\cite{Drake2018}.
X-ray probing is required to investigate the dense interiors of any samples and any measurements must be made in an ultrashort time frame due to its transient nature and ultrafast dynamics.
Because of these difficulties, many HED properties remain uncertain and are an on-going topic of research.
This includes equilibration rates~\cite{Cho2011}, opacities~\cite{Bailey2015,Kettle2016}, equations-of-state~\cite{Renaudin2013} and effects such as continuum lowering~\cite{Hoarty2013,Ciricosta2016} or non-thermal melting~\cite{Hartley2019}.
Understanding these properties is important, for example, for direct and indirect drive fusion experiments~\cite{Craxton2015,Lindl1995} as well as understanding the internal structure and evolution of large astrophysical objects~\cite{Remington2005}, including that of Earth itself~\cite{Nguyen2004,White2013}.
X-ray scattering techniques have been very successful in gaining information~\cite{GlenzerRedmer2009}, but provide limited access to the ion temperature and structure without assuming the sample is in thermodynamic equilibrium, or knowing the ionisation level, Debye temperature or ion-ion structure factor. %~\cite{Fletcher2015}
The cross-sections for X-ray scattering are also quite low, requiring an especially high brightness source such as an X-ray Free Electron Laser (XFEL) facility.

%-------------------------
% GENERAL AND LWFA SETUP FIGURE
\begin{figure*}[ht]
\includegraphics[width=0.9\textwidth]{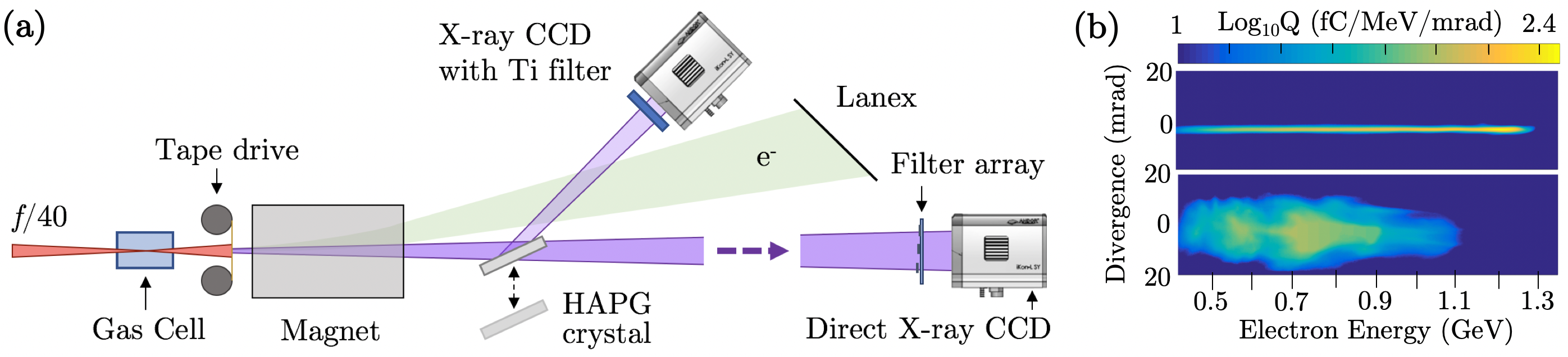}
\caption{\label{fig:setup-lwfa}(a) Experiment setup.
The LWFA X-rays can be measured on-axis or with a high resolution crystal spectrometer.
(b) Electron spectra where for the first stage of the gas cell,  $n_e=1.2\times10^{18}\textrm{cm}^{-3}$ (top) and $n_e=2.6\times10^{18}\textrm{cm}^{-3}$ (bottom).}
\end{figure*}
%-------------------------

X-ray absorption measurements can be used to understand both electron and ion dynamics on an atomic scale. %, allowing fundamental HED properties to be derived.
For example the absorption techniques of XANES (X-ray Absorption Near Edge Structure) and EXAFS (Extended X-ray Absorption Fine Structure) provide simultaneous measurements of the temperature and structure of the electrons and ions in a sample, as well as the ionisation state and more~\cite{Rehr2000,Dorchies2016}.
The ability to perform X-ray absorption measurements on an ultrafast timescale would enable a significant increase in the fundamental properties that can be derived from experiments involving HED samples, and other ultrafast phenomena. 
However, for HED experiments, single-shot measurements are crucial as many of the samples require a large amount of drive energy or complex target designs, making high repetition rates problematic.
It is also vital that multi-keV energies are available so that the inner shells of moderate-to-high Z elements can be probed, as well as large sample volumes.
Thus, for ultrafast X-ray absorption spectroscopy of HED samples to produce valuable and reliable data, a high-brightness smooth broadband spectrum X-ray probe in the multi-keV region is required.
The National Ignition Facility and the Omega Laser Facility have both been developing EXAFS diagnostics in the multi-keV regime~\cite{Krygier2018,Ping2013}. 
These sources however require 100's of joules of backlighter energy, and are over 100 picoseconds in duration.
In general, synchrotron facilities lack the required ultrashort pulse duration, and laser-plasma based approaches suffer additionally from low brightness and relatively noisy spectra~\cite{Mancic2010}.
XFELs have the required flux and pulse duration, but are monochromatic in nature.
Increased bandwidth techniques are being investigated~\cite{Prat2016}, however they lack a smooth broadband spectrum, making absorption measurements difficult.

A viable solution is to perform X-ray absorption measurements using a laser wakefield accelerator (LWFA).
These are the only currently available sources that provide bright bursts of broadband X-rays on the femtosecond time-scale~\cite{Phuoc2007} and their application in HED science has become an active research field.
To date however the source flux has required absorption spectra to be integrated over many shots or the photon energy range has been limited to lower energies (keV or less)~\cite{Mo2017, Smid2017, Mahieu2018}.
In this Letter we present the first single-shot multi-keV XANES measurement using the ultrashort X-rays from a LWFA source.
This was achieved by operating the LWFA in a tailored mode to generate high X-ray flux (more than 100 times that of previous measurements) and multi-keV photon energies, in tandem with an efficient and high-resolution single optic detector.
% This will allow...?

%-----------------------------------------------------------------------%
%   Experiment setup
%-----------------------------------------------------------------------%

The experiment was conducted using the Gemini Laser at the Central Laser Facility (U.K.). 
An overview of the experiment setup can be seen in Fig.~\ref{fig:setup-lwfa} (a)~\cite{OurData}.
The drive laser (800 nm) was focused using an f/40 geometry into a gas cell.
Each laser pulse (provided at 0.05 Hz) had a duration of $47\pm5$ fs and contained $9\pm0.3$ J. 
These pulses were focused to a spot of $50\pm2~\mu\textrm{m} \times 43\pm1~\mu\textrm{m}$ FWHM, with the central FWHM containing $43\pm2\%$ of the energy.
This provided an on-target intensity of $4.9 \pm 0.6 \times10^{18}~\textrm{W/cm}^2$ and an average laser strength parameter of $a_0\approx1.5$.
As the laser pulse traveled through the gas, it drove an LWFA~\cite{TajimaDawson1979}, where the electrons liberated from the atoms were expelled by the ponderomotive force of the laser, creating an ion cavity in its wake.
The strong electric field inside the cavity can subsequently accelerate electrons to gigaelectronvolt energies in just a few centimetres~\cite{Leemans2006, Kneip2009}.
Our LWFA operated using a two-stage gas cell~\cite{Pollock2011}~\cite{Hussein2019}. 
The first stage ($\textrm{3 mm}$ long) was filled with a 98\% He + 2\% $\textrm{N}_2$ gas mix, and the second stage (19.6 mm long) was filled with He. 
Electrons were injected in the first stage using ionization injection~\cite{Pak2010, McGuffey2010}. % RowlandsRees2008, 
The second stage provided the acceleration of the electrons. % to gigaelectron volt energies.
While in the back of the ion cavity the electrons perform betatron oscillations around the laser axis, producing high energy X-rays~\cite{Rousse2004, Kneip2010}.
The on-axis intensity spectrum is synchrotron-like and characterised by the critical energy $E_{\mathrm{crit}}$ and is given by $d^2I/(dEd\Omega) \propto (E/E_{\mathrm{crit}})^2 \mathcal{K}^2_{2/3}[E/(2E_{\mathrm{crit}})]$, where $\mathcal{K}_{2/3}[x]$ is a modified Bessel function of order $2/3$.
The X-ray pulse emission is of similar duration to that of the electron bunch which is typically on the order of 10 femtoseconds~\cite{Lundh2011}.
The source size is on the order of microns and the emission is directed in a tight cone along the propagation axis, with a divergence of $\lesssim 20~\textrm{mrad}$ FWHM.
%The source size is on the order of a micron, perfect for high resolution imaging.

\begin{figure}
\includegraphics[width=0.48\textwidth]{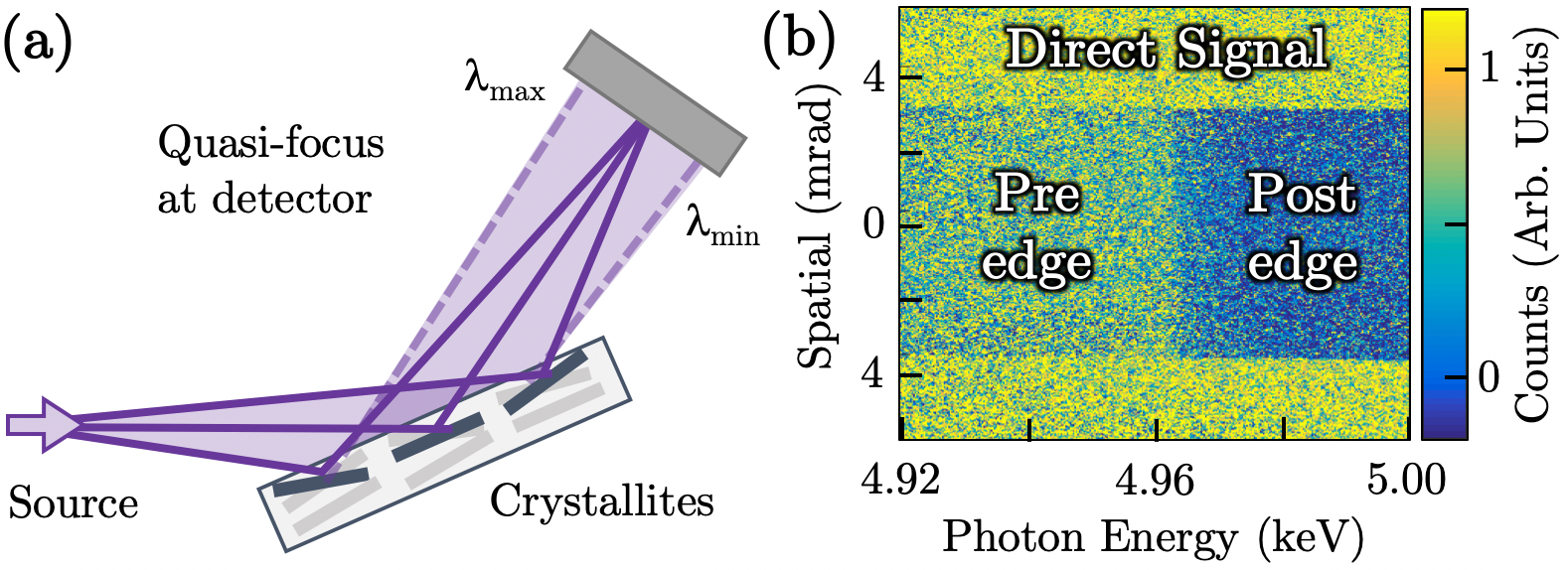}
\caption{\label{fig:setup-xanes}(a) Illustration of mosaic quasi-focusing.
Solid rays depict a single wavelength focused at the detector.
Dashed rays indicate the minimum/maximum wavelength (dictated by the source divergence).
(b) Single-shot crystal spectrometer image corrected for the crystal reflectivity profile.}
\end{figure}

After the X-rays exit the accelerator, a replenishable tape drive was used to dump the remaining laser energy, and a high strength magnet ($\approx0.8$ T, 10 cm) was used as an electron energy spectrometer. 
The tape is made of polyimide plastic and is $25~\mu\textrm{m}$ thick.
It has a transmission of over $90\%$ for X-ray energies over 5 keV.
Two example electron spectra can be seen in Fig.~\ref{fig:setup-lwfa} (b).
The X-ray spectrum was measured with high-resolution using the reflection from a crystal (protected from laser damage by the sacrificial tape drive) or directly imaged through a set of metallic filters to estimate the broadband spectrum~\cite{Kneip2010}.
%An example image from a region of the direct CCD can be seen in Fig.~\ref{fig:setup} (c).
The high resolution spectral measurements of the X-rays were made over a range of $\approx~80$ eV.
A $100~\mu\textrm{m}$ thick HAPG (Highly Annealed Pyrolytic Graphite) crystal with $\approx0.1^\circ$ mosaic spread on a $2\times6$ cm BK7 substrate was used.
Mosaic crystals (as opposed to perfect crystals) are made up of many smaller crystallite planes that have a random nature to their orientation.
The angles of these planes are seen to have a normal distribution with a width of less than a degree.
However, this spread in crystallite angles throughout the crystal structure is responsible for increasing the reflection efficiency, as the Bragg condition for any given photon wavelength can now be satisfied over a larger surface area of the crystal.
This effect is known as quasi-focusing and provides reflection efficiencies of over 10 times that of a perfect crystal, while maintaining the high-spectral resolution if used in a 1-to-1 geometry~\cite{Legall2009}. 
See Fig.~\ref{fig:setup-xanes} (a) for an illustration of this operation.
The source to crystal (and crystal to detector) distance was $41\pm1~\textrm{cm}$.
Ray-tracing simulations estimate the instrument function of this spectrometer to have a width of $\approx2.2$ eV, consistent with the estimates of Zastrau \textit{et al.}~\cite{Zastrau2013}.
This resolution is dominated by quasi-focusing of the mosaic crystal, as opposed to the single crystal plane broadening or source broadening effects (as the LWFA source is on the order of microns).
By using a single high-reflectivity optic we have optimised the overall efficiency of the X-ray detector while maintaining high-spectral resolution.
A $10~\mu\textrm{m}$ thick titanium sample strip was placed in front of the CCD to record absorption features around its K-edge (4966 eV). 

%However, due to this inherent angular spread of Bragg conditions, the spectral content reflected from the crystal at any given point is widened, resulting in a reduced spectral resolution at the detector.
%In our case we have used a (nominally) $100~\mu \textrm{m}$ coating of $0.1^\circ$ mosaicity HAPG (Highly Annealed Pyroltyic Graphite) on a 2x6 cm BK7 substrate.
%Ray tracing simulations show a 2.2 eV spectral resolution resulting from the mosaic spread.
%This crystal differs from the HOPG (Highly Ordered Pyroltyic Graphite) used by Smid et al., which has a $0.8^\circ$ mosaicity providing an even higher reflection efficiency, but at the expense of a reduced spectral resolution (5.5 eV).

%-----------------------------------------------------------------------%
%   Results and Analysis
%-----------------------------------------------------------------------%

% ------------------
% LWFA Overview
% ------------------
For plasma densities of $n_e=1.2\times10^{18}\textrm{cm}^{-3}$ and $n_e=2.3\times10^{18}\textrm{cm}^{-3}$ in the first and second cell stages respectively, electron beams with a maximum energy at 1.2~GeV and a divergence of 1~mrad were observed on the magnetic spectrometer (see Fig.~\ref{fig:setup-lwfa} (b) top).
However, we found that we were able to generate ten times more X-ray flux by increasing the plasma density to $2.6\times10^{18}\textrm{cm}^{-3}$ in both stages.
At this density, the observed electron beam had a lower maximum energy, but a greater total charge and transverse momentum (see Fig.~\ref{fig:setup-lwfa} (b) bottom).
This also increased the X-ray divergence to $\approx 15~\textrm{mrad}$ FWHM.
As the divergence of the X-ray source provides the range of different incident angles upon the crystal spectrometer and the spectral spread of the detector is achieved by satisfying the Bragg condition at different angles, a more divergent beam leads to a wider accessible spectral range.
For the high flux shots the direct filter pack measured a mean critical energy of $E_{\rm{crit}}=9.9\pm1.5$~keV, and the entire beam contained $7.2\pm2.8\times10^5$ photons/eV at 5 keV, 
%equating to a brightness of nearly $10^{23}~\textrm{photons/s/mm}^2/\textrm{mrad}^2/0.1\% \textrm{BW}$, 
comparable to the highest X-ray flux observed in previous LWFA measurements~\cite{Cole2015,Hussein2019}. %Cole2015,
The shot-to-shot standard deviation here is combined with the systematic errors in quadrature.

%-------------------------
% CRYSTAL SETUP AND RAW XANES FIGURE
%-------------------------

% ------------------
% Data Processing
% ------------------
A single-shot image from the crystal spectrometer can be seen in Fig.~\ref{fig:setup-xanes} (b).
It has been background corrected, and the spatial variations in the signal due to the mosaic crystal structure have been folded out (see the supplemental material for further details).
The horizontal axis corresponds to the X-ray energy, while the vertical axis provides spatial information perpendicular to the dispersion direction.
The shadow of the titanium sample foil along the central region provides the absorption profile around the inner K-shell, whereas the direct signal either side measures the X-ray yield and smoothness.

In the direct signal region for the brightest shot, we measure $1.2\pm0.2\times10^6$ photons/eV. % in the 5 keV region. 
Assuming a Poisson distribution, the random statistical noise should be $\sqrt{N_{ph}}$ where $N_{ph}$ is number of photons, i.e we should have a signal-to-noise level of $\approx$~1100:1 per eV.
Our direct signal exhibits a signal-to-noise of 300:1 (standard deviation in the photon yield per eV, $0.34\%$ of the signal level). 
One of the main contributions to the noise comes from an underlying background that is combined with the X-ray signal from the crystal reflection.
This noise is present even on shots where the X-ray crystal (but not the CCD) was removed from the beamline, indicating that the source is not inherent to the measurement.
The background is seen to scale linearly with the total charge of the electron beam.
Single-hit photon analysis of low-charge shots also suggest that the CCD hits are from a broad spectra of hard X-rays and occasional high energy particle hits.
This is consistent with the accelerated electrons interacting with the target chamber and creating secondary particles which produce the background noise.
The measured standard deviation noise on an X-ray shot (with the crystal in place) is found to be on average $\lesssim 12\%$ higher than that of a background shot (without the crystal).
Assuming the noise sources add in quadrature, this suggests the statistical noise inherent in the betatron signal is less than half of the electron-beam produced background noise. 
$\sigma_{\rm{signal}}=\sigma_{\rm{bg}}\sqrt{(\sigma_{\rm{all}}/\sigma_{\rm{bg}})^2 - 1}$.
Importantly, it should therefore be possible to significantly reduce the background with improved shielding and appropriate electron beam dumping.

\begin{figure}
\includegraphics[width=0.45\textwidth]{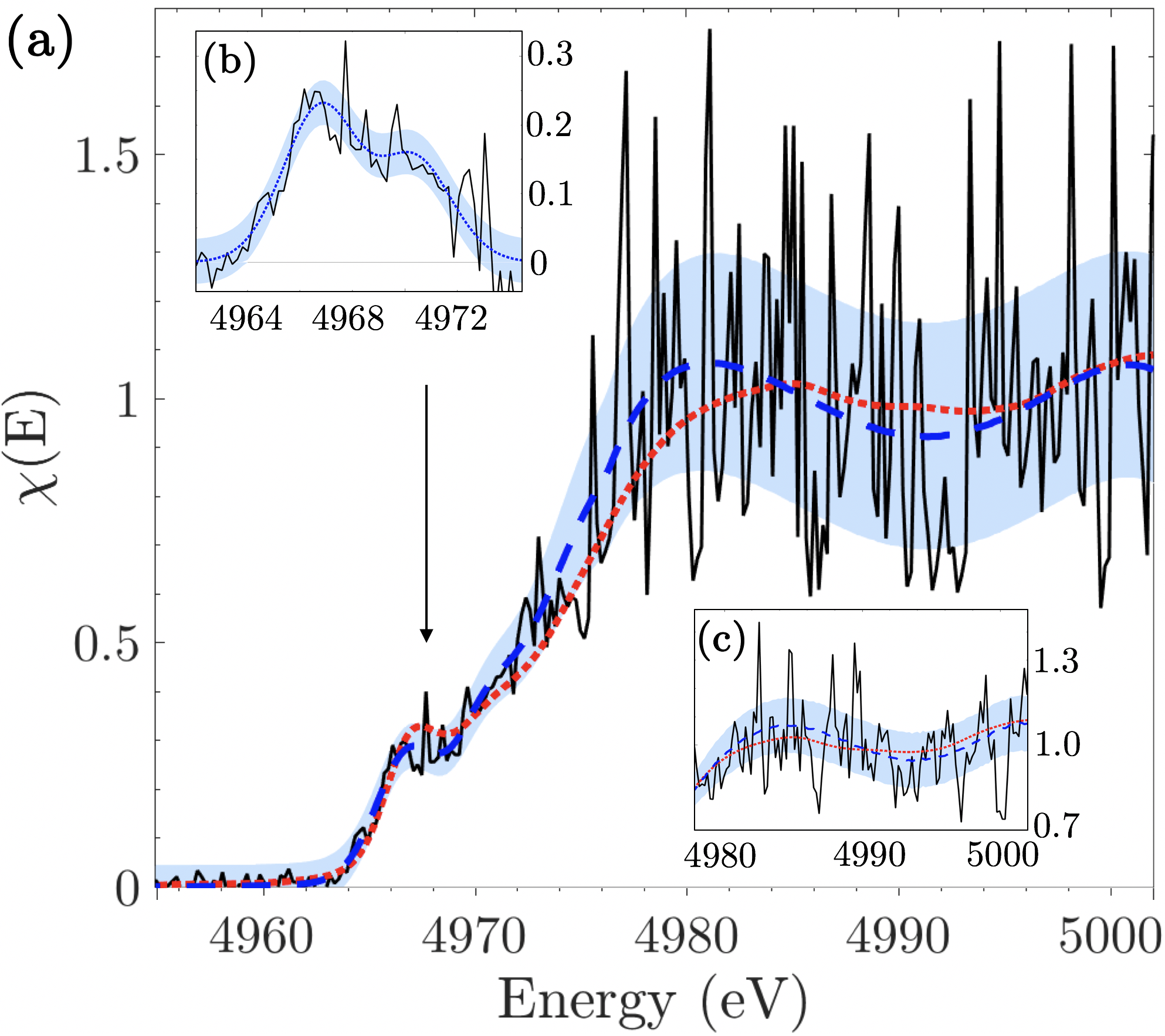}
\caption{\label{fig:singleshot} (a) Single-shot normalised absorption data (solid black) compared to a synchrotron reference measurement~\cite{Bleith2013} (dotted red). 
The fitted profile for our data is given (dashed blue) with the light blue area indicating the measurement error. 
%See text for more details.
(b) Double gaussian fit to the pre-edge features. 
%, after the data in (a) has been ``flattened'' by subtracting the sigmoid fit.
(c) The same result as in (a), but averaged over 11 shots.}
\end{figure}

%An averaged lineout is taken across the spatial width of the titanium sample.
Fig.~\ref{fig:singleshot} depicts the measured absorption profile for a single shot (solid black).
It is compared alongside reference data for titanium taken previously at a synchrotron facility~\cite{Bleith2013} (dotted red).
To facilitate the comparison we have used a standard XANES procedure for normalising the profile~\cite{Newville2014}.
%A polynomial $\mu_0(E)$ is fitted to the signal before the K-edge, and another $\mu_1(E)$ to after the edge.
%We subtract the previous from the absorption profile $\mu(E)$, then divide by $\mu_1(E)$, to give the normalised absorption profile $\chi(E) = \frac{\mu(E) - \mu_0(E)}{\mu_1(E)}$.
This has the added benefit of not requiring a direct spectrum (no sample absorption) to be measured, as long as the signal is relatively smooth and stable (a key strength of the betatron radiation from the LWFA).
The reference data, which already has an inherent instrument width, has had a 2 eV FWHM instrument function applied (to match our detector resolution).

% ------------------
% SINGLE SHOT FIGURE
% ------------------

% ------------------
% Single-shot results
% ------------------

The single-shot measurement provides a clear match to the rising slope structure ($<4970~\textrm{eV}$) as well as emulating the pre-edge feature at 4967 eV.
This pre-edge feature is a set of forbidden transitions into the 3d shell, allowed by 3d-4p mixing, and provides information about the bonding properties of the sample~\cite{Dorchies2016}.
The underlying slope of the edge corresponds to the density of free states, and provides the temperature of the electron distribution (which was 300 K in our case). 

To examine the profile we fit various structures to the different aspects of its shape.
First we fit a sigmoid function to replicate the Fermi distribution of the electrons, where the width is proportional to the temperature.
%$1-f(E)=A/(1+e^{(E-E_0)/C})$, where $E_0$ is the photon energy at the midpoint, $A$ is the magnitude, and $C$ dictates the width of the distribution.  
%The value of $C$ is proportional to the temperature $kT$, having to account for the broadening due to the instrument function of the spectrometer.
% The fitted sigmoid $C$ value agreed that of a 0.025 eV room temperature measurement (within the error bar on the instrument function of $\pm0.2~\rm{eV}$).
%Any increase in the electron temperature will further broaden this function.
%After subtracting this sigmoid fit, we are left with a ``flattened'' profile (see supplemental information for a depiction).
We fit a double gaussian to the two pre-edge forbidden transitions, and a fifth-order polynomial to the oscillatory component of the XANES interference features after the edge.
Fig.~\ref{fig:singleshot} (b) illustrates the double gaussian fit after the sigmoid subtraction.
These fit components are combined and an error bar equal to the standard error of the fit combined with the error in the crystal reflectivity is added. % in quadrature
See Fig.~\ref{fig:singleshot}, dashed blue line and grey shaded areas respectively.
This fitting procedure allows us to quantify how well our data agrees with the reference data.

We assess our resolution on a single shot, by studying the fitting fluctuations over several consecutive shots.
% Over eleven shots we managed to fit
We observe a fit to the edge position with a standard error of 0.17 eV. 
The standard deviation in the position of the foot ($10\%$ value) before the pre-edge features is 0.28 eV.
Assuming a Fermi distribution of the electrons we estimate this would allow a resolvable change in electron temperature of $\approx0.4~\textrm{eV}$ on a single shot.
%this would allow a change in electron temperature of $\approx0.4~\textrm{eV}$ to be resolvable.
The amplitude of the pre-edge gaussians have an 18\% error.
In summary, on a single-shot measurement we are capable of quantitatively resolving electronic structure information and electron temperature with sub-eV accuracy.

% ------------------
% post-edge estimations
% ------------------
The post edge modulations in the profile also contain valuable information regarding the ion component of the sample.
It has been estimated that a signal-to-noise of 1000:1 is required to make a high quality EXAFS measurements of the ion peak beyond the edge, with good statistics~\cite{Albert2014}.
%In an effort to emulate an improved signal-to-noise ratio for our data, we perform the analysis on an integration of 11 shots.
We can emulate the expected improvements to the signal-to-noise that will be achieved with improved electron beam shielding by averaging the data over 11 shots.
The inset of Fig.~\ref{fig:singleshot} depicts the measured absorption profile (solid black), our resulting fit (blue dashed) with shaded error bars, and the synchrotron reference (red dotted)~\cite{Bleith2013}.
The error magnitude in the signal region post-edge has been reduced by a factor of two.
From the noise discussion before it was seen that the background noise present in our data contributes at least twice that of the X-ray signal from the crystal reflection.
Therefore, a signal-to-noise similar to the integrated shots should be achieved (or bettered) for a single-shot with an improved electron beam dump and detector shielding.
The contrast in the absorption profile can also be improved by a factor of two by choosing an optimal sample thickness ($\textrm{1}/e$ absorption depth).
From comparison to a range of density functional theory (DFT) simulations at different ion temperatures, the resolution achieved in the post-edge modulation structure (assuming noise reduction) should be sufficient to see a change of $\approx0.5~\textrm{eV}$ in ion temperature (via the ``flattening" of the modulation structure). 
This is extremely valuable information, especially in tandem with the electron temperature accessed via the absorption edge slope. 

% ------------------
%   DFT sims
% ------------------

% ------------------
% DFT Figure
\begin{figure}
\includegraphics[width=0.45\textwidth]{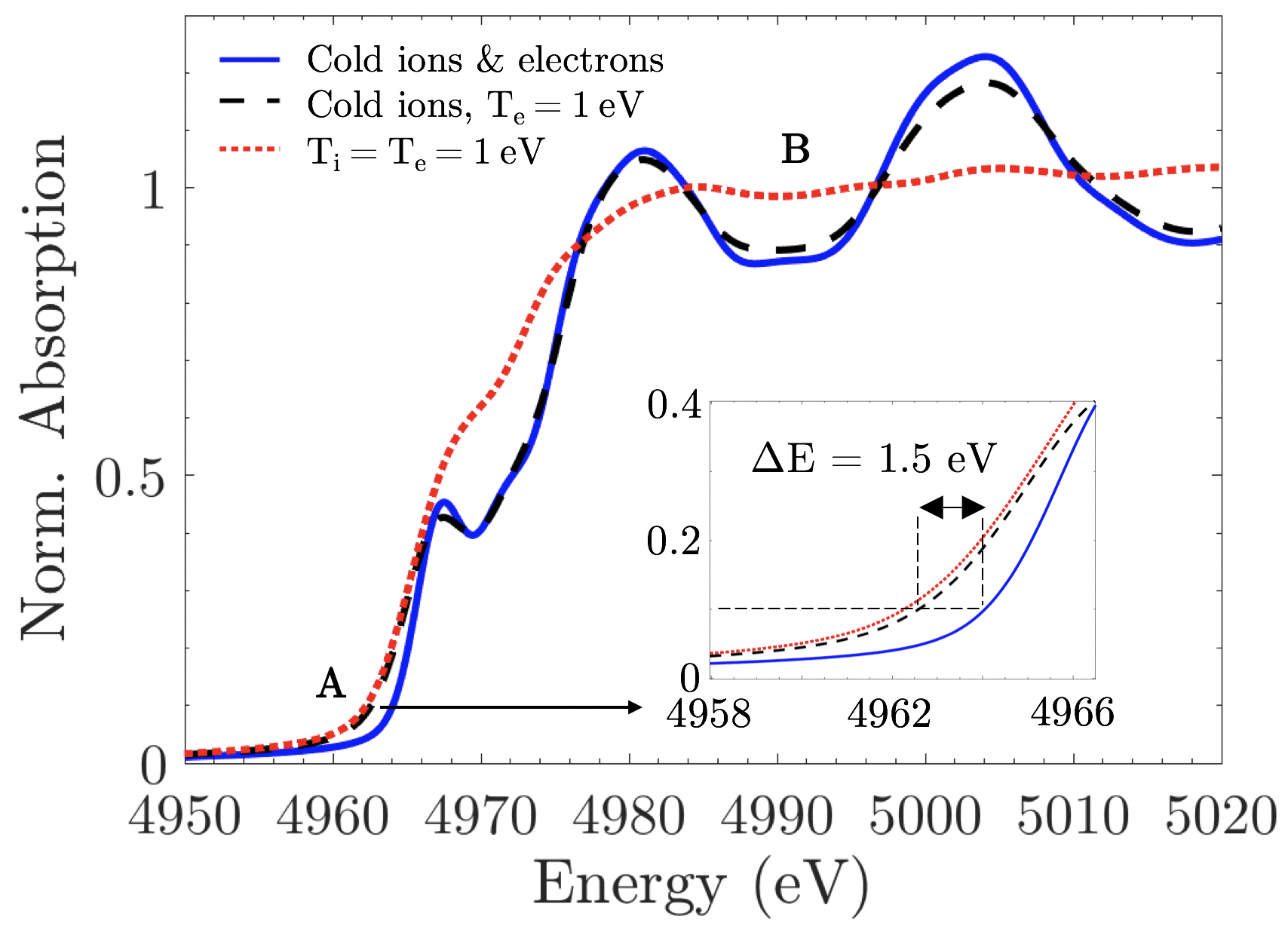}
\caption{\label{fig:DFT}DFT simulations for titanium at various heating conditions (room temperature, non-equilibrated and $T=1~\textrm{eV}$).}
\end{figure}
% ------------------

With that in mind, we discuss the possibility of investigating a non-equilibrated HED sample.
Fig.~\ref{fig:DFT} depicts DFT results for titanium using \textit{GPAW}~\cite{Enkovaara2010}.
% also added some Lorentzian width to simulate the lifetime broadening
A $3~\textrm{eV}$ instrument function is applied to (conservatively) emulate experimental measurements.
The normalised absorption profile is given for three scenarios; a cold foil (solid blue), a foil in a non-equilibrated situation with $T_e=1~\textrm{eV}$ and cold ions (dashed black), and finally a foil with $T_e=T_i=1~\textrm{eV}$ (dotted red).
The effect on the pre-edge profile due to the increase in electron temperature can be clearly seen in the latter two cases with a $1.5~\rm{eV}$ shift in the position of the foot.
This is highlighted in Fig.~\ref{fig:DFT} with the $A$ marker and the zoomed inset.
Our single-shot data is more than capable of resolving such a change in the profile.
The loss of the post-edge modulations is only evident in the final case (highlighted by $B$), when the ion temperature is increased and the radial distribution function flattens.
For any single measured absorption profile it would be possible to independently deduce the electronic and ionic structure and temperature using these independent shifts in the absorption structure.
This would for example allow the electron-ion equilibration rate to be directly diagnosed, a highly sought after measurement~\cite{Cho2011}.

With more high-intensity lasers coming online around the world that can generate the required LWFA X-ray source, especially in tandem with sample drivers such as XFELs or high-energy laser systems, single-shot X-ray absorption spectroscopy offers great capabilities in making ultrafast measurements of many fundamental processes in HED science.
This includes the use of XANES for electron-ion equilibration rate measurements as described above, but also high-quality ultrafast X-ray absorption data could be used to investigate continuum lowering effects~\cite{Hoarty2013} or informing solar opacity debates~\cite{Bailey2015}.
Commercial and industrial applications of multi-keV absorption also become feasible.
For example the ultrafast phase-changes that can be measured via the EXAFS structures are important in battery~\cite{Wang2016} and memory storage~\cite{Zalden2019} studies.
Single-shot measurements would increase the efficiency of such studies and the femtosecond pulse length would resolve the phase dynamics further.
Finally, development of the LWFA X-ray properties is an on-going and active research area.
Any advances will extend the capabilities of this technique.
For example an increased X-ray flux will allow higher spectral resolution measurements by allowing less efficient perfect crystals to be used in the X-ray spectrometer.

%-----------------------------------------------------------------------%
%   Conclusions
%-----------------------------------------------------------------------%

In conclusion, we have demonstrated a novel experimental technique for performing single-shot ultrafast X-ray absorption measurements of the electronic structure of a sample.
These measurements were made possible using the smooth broadband multi-keV synchrotron radiation produced by a laser-driven plasma wakefield accelerator.
To achieve a single-shot measurement we generated a high-energy and high-charge electron beam from the accelerator to produce an X-ray flux of $>10^6$ photons/eV and implemented a suitable high-resolution and high-efficiency X-ray spectrometer using a single reflection optic.
We were able to perform XANES measurements of room temperature titanium, with the results in agreement with measurements taken at a 3rd generation light source.
Our data provides not only the electron temperature distribution %via the slope 
but information on any additionally supported pre-edge transitions.
With minor improvements to our experimental set up %(mainly background noise removal),
one should be able to access the ion component further from the absorption edge and make ultrafast single-shot EXAFS measurements of mid-to-high Z elements.
This will allow the simultaneous measurement of the electronic and ionic temperature and structure of high-energy-density samples on a timescale of tens of femtoseconds, making significant new areas of research possible.

We wish to acknowledge the support of the staff at the Central Laser Facility.
This project has received funding from the European Research Council (ERC) under the European Union's Horizon 2020 research and innovation programme (grant agreement no 682399), as well as the U.S. Department of Energy Office for Fusion Energy Sciences, project DE-SC0019186, and the Knut and Alice Wallenberg Foundation.

%\end{acknowledgments}

%-----------------------------------------------------------------------%
\bibliography{main} % Produces the bibliography via BibTeX.
%\bibliographystyle{plain}
%-----------------------------------------------------------------------%
\end{document}

\title{Supplemental Material for\\ ``Single-shot multi-keV X-ray absorption spectroscopy using an ultrashort laser wakefield accelerator source''}
%setting up commands for each affiliation
\newcommand{\JAI}
{The John Adams Institute for Accelerator Science, Imperial College London, London, SW7 2AZ, UK}
\newcommand{\GOLP}
{GoLP/Instituto de Plasmas e Fus\~{a}o Nuclear, Instituto Superior T\'{e}cnico, U.L., Lisboa 1049-001, Portugal}
\newcommand{\CUOS}
{Center for Ultrafast Optical Science, University of Michigan, Ann Arbor, Michigan 48109-2099, USA}
\newcommand{\Lund}
{Department of Physics, Lund University, P.O. Box 118, S-22100, Lund, Sweden}
\newcommand{\Lancaster}{Physics Department, Lancaster University, Lancaster LA1 4YB, United Kingdom}
\newcommand{\CLF}{Central Laser Facility, STFC Rutherford Appleton Laboratory, Didcot OX11 0QX, UK}
\newcommand{\HZDR}{Helmholtz-Zentrum Dresden-Rossendorf, Bautzner Landstrasse 400, 01328 Dresden, Germany}
\newcommand{\ASCR}{Institute of Physics of the ASCR, Na Slovance 1999/2, 182 21 Prague, Czech Republic}
\newcommand{\TUD}{Technische Universit{\"a}t Dresden, D-01069 Dresden, Germany}
\newcommand{\LLNL}{Lawrence Livermore National Laboratory (LLNL), Livermore, California 94550, USA}

\author{B.~Kettle}
\email{b.kettle@imperial.ac.uk}
    %\homepage{http://www.Second.institution.edu/~Charlie.Author.}
    %\altaffiliation[Also at ]{Physics Department, XYZ University.}%Lines break automatically or can be forced with 
\affiliation{\JAI}

\author{E.~Gerstmayr}
\affiliation{\JAI}

\author{M.J.V.~Streeter}
\altaffiliation[Current address: ]{\JAI}
\affiliation{\Lancaster}

%alphabetical from here?
\author{F.~Albert}
\affiliation{\LLNL}    

\author{R.A.~Baggott}
\affiliation{\JAI}

\author{N.~Bourgeois}
\affiliation{\CLF}

\author{J.M.~Cole}
\affiliation{\JAI}

\author{S.~Dann} 
\altaffiliation[Current address: ]{\CLF}
\affiliation{\Lancaster}

\author{K.~Falk}
\affiliation{\HZDR}
\affiliation{\TUD}
\affiliation{\ASCR}

\author{I.~Gallardo Gonz\'{a}lez}
\affiliation{\Lund}

\author{A.E.~Hussein}
\affiliation{\CUOS}

\author{N.~Lemos}
\affiliation{\LLNL}

\author{N.C.~Lopes}
\affiliation{\GOLP}

\author{O.~Lundh}
\affiliation{\Lund}

\author{Y.~Ma}
\altaffiliation[Current address: ]{\CUOS}
\affiliation{\Lancaster}

\author{S.J.~Rose}
\affiliation{\JAI}

\author{C.~Spindloe}
\affiliation{\CLF}

\author{D.R.~Symes}
\affiliation{\CLF}

\author{M.~\v{S}m\'{i}d}
\affiliation{\HZDR}

\author{A.G.R.~Thomas}
\affiliation{\CUOS}
\affiliation{\Lancaster}

\author{R.~Watt}
\affiliation{\JAI}

\author{S.P.D.~Mangles} 
\affiliation{\JAI}
\email{stuart.mangles@imperial.ac.uk}

%\date{\today}
\maketitle
%--------------------------------------------------------------------------------------%

In this Supplemental Material we present technical details related to the data processing of the X-ray diagnostics in the main text.

%--------------------------------------------------------------------------------------%
% Syncrotron spectrum fitting
%--------------------------------------------------------------------------------------%
\section{Direct filter pack measurements of the X-ray spectrum}

\begin{figure}[ht]
\includegraphics[width=0.2\textwidth]{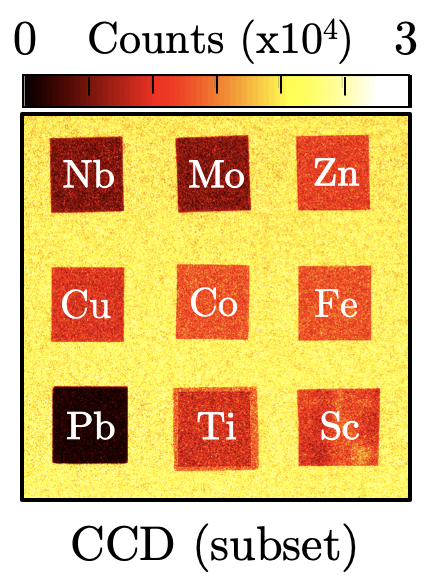}
\caption{\label{fig:filterpack} On-axis broadband spectral measurement using filter array transmissions.}
\end{figure}

An example image from the on-axis broadband spectral measurement of the X-rays can be seen in Fig.~\ref{fig:filterpack}.
As the X-ray transmission through each filter element is spectrally dependent, a synchrotron spectrum can be fitted to the relative transmission through all the elements.
We vary the critical energy $E_{\rm{crit}}$ of the spectrum to minimise the sum of the squared differences from the filter transmissions.
Seven different elements where used in total; Ti, Fe, Co, Cu, Zn, Nb, and Mo with an additional Pb element with zero transmission.
The elements have thicknesses of (respectively); 17.3, 5.6, 5.4, 9.2, 10, 24.5 and 20 microns.
The error in each thickness measurement is $\pm0.3~\mu\textrm{m}$.
A scandium (Sc) filter was also present, but found to have an unreliable thickness across the segment and was hence not used for the spectral retrieval.

%--------------------------------------------------------------------------------------%
% XANES corrections
%--------------------------------------------------------------------------------------%
\section{Correcting for variable crystal reflectivity}

\begin{figure}
\includegraphics[width=0.98\textwidth]{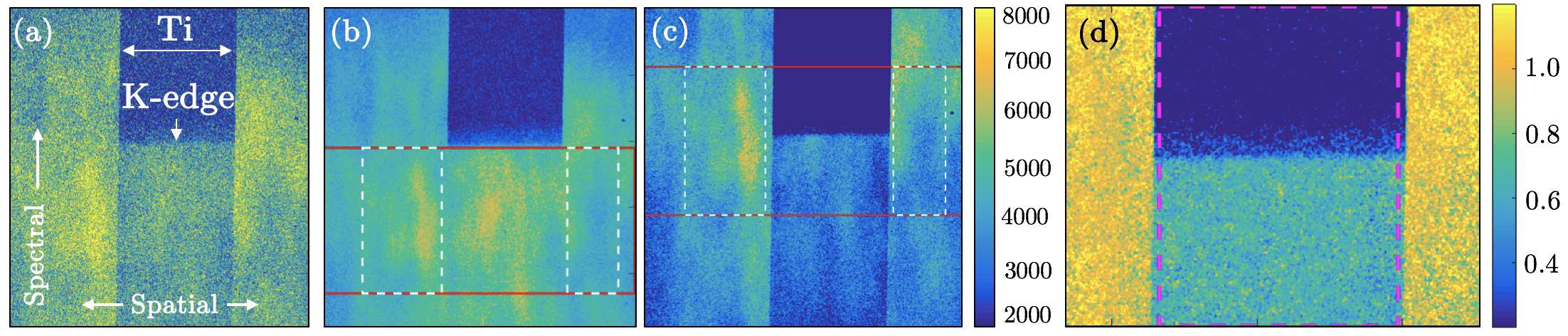}
\caption{\label{fig:data_proc}(a) Single-shot crystal spectrometer image. (b) 19 integrated shots, $\theta=21.85^\circ$. (c) 11 integrated shots, $\theta=22.05^\circ$. (d) The data of (c) once corrected for variations and normalised.}
\end{figure}

An example raw single-shot image from the crystal spectrometer can be seen in Fig.~\ref{fig:data_proc} (a).
The horizontal axis provides spatial information, while the X-ray photons are spectrally spread along the vertical axis.
Lighter colours indicates higher signal.
The shadow of the titanium sample can clearly be seen in the center of the image.
The cut in absorption due to the K-edge is also highlighted.
Although the signal is apparent, it varies in intensity across the image.
The majority of this variation is not inherent in the X-ray source emission, but is an artifact of the reflection from the mosaic crystal surface (and is repeated on every shot).
It is possible to measure this crystal structure directly (without any sample in place), but this was not performed on this campaign.
In order to correct for the signal variation during our experiment, measurements were taken at two different crystal angle positions.
By changing the crystal angle and keeping the CCD constant, the spectral spread of the detector does not change, but the relative location of the signal shifts.
Fig.~\ref{fig:data_proc} (b) and (c) depict the integrated images from crystal angles of $21.85^\circ$ (19 shots) and $22.05^\circ$ (11 shots) respectively.
The position of the K-edge feature remains fixed, but the structure of the signal shifts vertically.
A 2D map is fitted to the signal structure before the K-edge in Fig.~\ref{fig:data_proc} (b) (indicated by the solid red box), and this map is used in Fig.~\ref{fig:data_proc} (c) to estimate the original signal structure, before absorption, for photon energies above the K-edge.
The position and relative intensity of this correction map is set by the reference areas either side of the sample (indicated by the dashed white lines).
A signal variation corrected image can be seen in Fig.~\ref{fig:data_proc} (d).
The relative error between the two signals in the known reference regions is $\approx 1\%$, and this is assumed to be the same in the absorption measurement areas.

%--------------------------------------------------------------------------------------%
% Background Noise
%--------------------------------------------------------------------------------------%
\section{Background Noise}

\begin{figure}[h]
\includegraphics[width=0.7\textwidth]{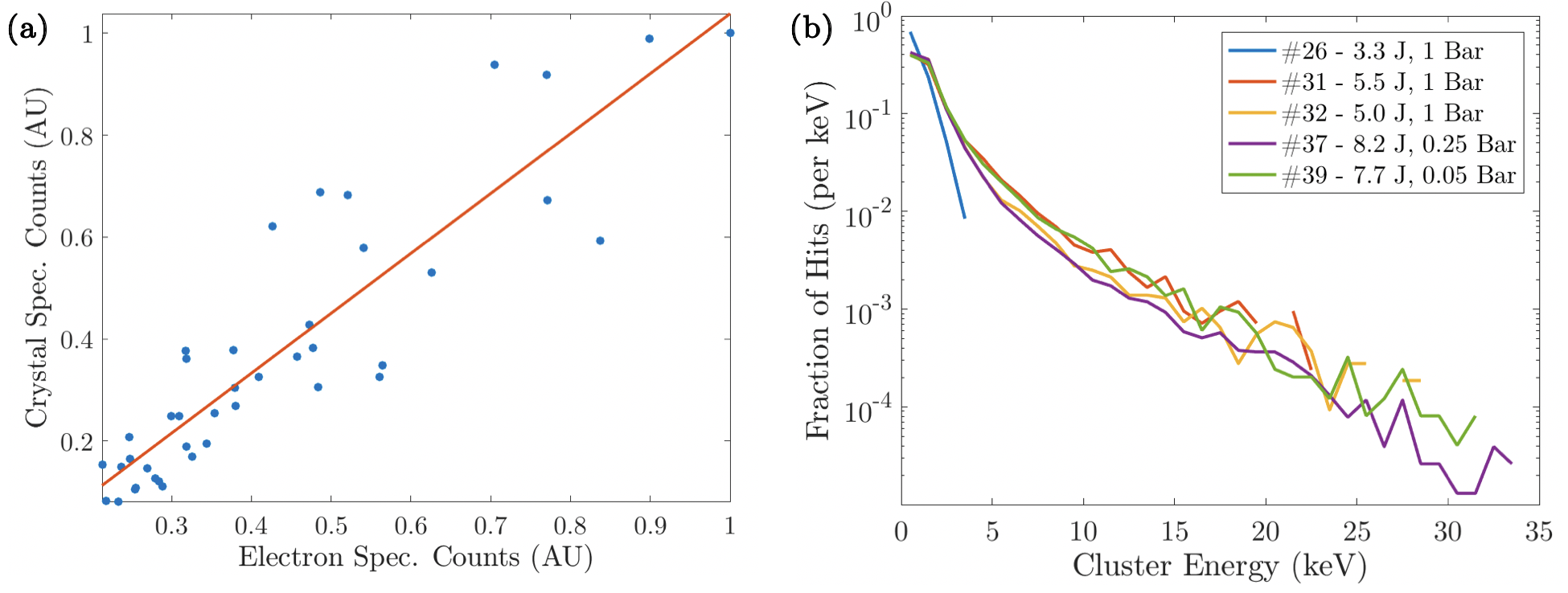}
\caption{\label{fig:bg}(a) Crystal spectrometer CCD counts as a function of measured electron charge for shots with no crystal in place. (b) Single-hit cluster energies for various background shots at low background flux.}
\end{figure}

We assess the background level on the crystal spectrometer by summing the CCD counts on shots where the crystal is moved out of the direct path of the X-ray beam, and hence no signal is being reflected towards the detector.
For each of these shots we also sum the total electron charge simultaneously detected on the electron spectrometer.
Fig.~\ref{fig:bg} (a) depicts a linear correlation for the number of CCD counts as a function of electron charge.

In an attempt to further characterise this background signal we perform a single-hit cluster analysis of the individual CCD strikes on shots with a low enough level of flux (thus avoiding strike pile-up).
Less than $3\%$ of CCD pixels registered a value above threshold for all shots used in the single-hit analysis.
A cluster combining algorithm was used to calculate the total freed energy for each individual identifiable strike on the CCD.
Fig.~\ref{fig:bg} (b) depicts the number of strikes as a function of total contained energy in each cluster.
For the shots depicted, various plasma densities and laser energies were used, corresponding to varying levels of electron charge being generated from the LWFA.
Shot 26 in the blue has no detectable electron charge driven, and represents the dark current of the CCD.
All shots apart from 26 show a broadband spectra of cluster energies tailing out to 30 keV (which is also where the quantum efficiency of the CCD falls off).

These results indicate that in the current configuration the background noise is being produced by the accelerated electrons interacting with the target chamber and creating secondary noise sources.
Importantly, it should therefore be possible to significantly reduce the background with improved shielding and appropriate electron beam dumping, away from the CCD.

%--------------------------------------------------------------------------------------%
% X-ray absorption profile normalisation
%--------------------------------------------------------------------------------------%
\section{X-ray absorption profile normalisation}

We have used a standard XANES/EXAFS procedure for normalising the absorption profile~\cite{Newville2014}.
A polynomial $\mu_0(E)$ is fitted to the signal before the K-edge, and another $\mu_1(E)$ to after the edge.
We subtract the previous from the absorption profile $\mu(E)$, then divide by $\mu_1(E)$, to give the normalised absorption profile $\chi(E) = \frac{\mu(E) - \mu_0(E)}{\mu_1(E)}$.

%--------------------------------------------------------------------------------------%
% Fitting Procedure
%--------------------------------------------------------------------------------------%
\section{X-ray absorption profile fitting procedure}

\begin{figure}[h]
\includegraphics[width=0.7\textwidth]{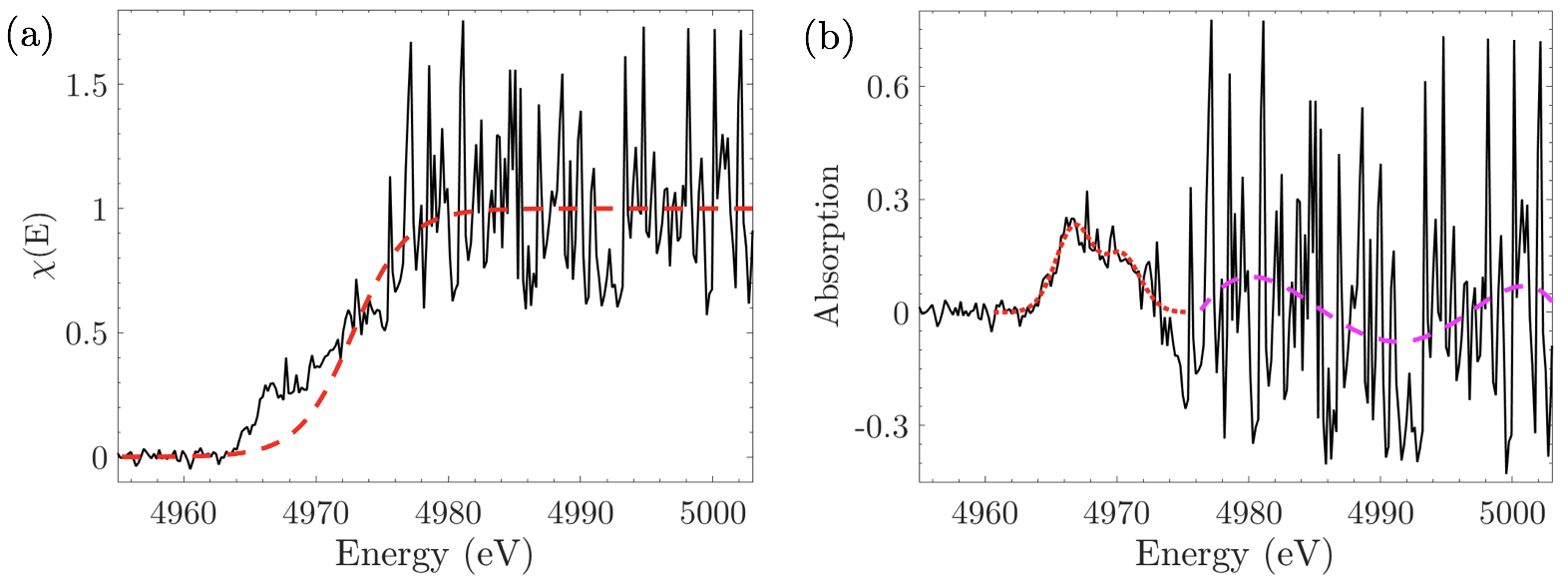}
\caption{\label{fig:fitting_proc}(a) Sigmoid fitting and (b) gaussian and oscillatory feature fitting.}
\end{figure}

To examine the normalised absorption profile, we fit various structures to the different aspects of its shape.
First we fit a sigmoid function to replicate the Fermi function for the electron distribution.
See Fig.~\ref{fig:fitting_proc} (a) for an example.
$1-f(E)=A/(1+e^{(E-E_0)/C})$, where $E_0$ is the photon energy at the midpoint, $A$ is the magnitude, and $C$ dictates the width of the distribution.  
The value of $C$ is proportional to the temperature $kT$, having to account for the broadening due to the instrument function of the spectrometer.
A least-squares fit was used to match the slope above the pre-edge feature ($E > 4973~\textrm{eV}$), while $E_0$ and $C$ are varied.
The fitted sigmoid $C$ value agreed with that of a 0.025 eV room temperature measurement (within the error bar on the instrument function of $\pm0.2~\rm{eV}$).
Any increase in the electron temperature will broaden this function with an increase in the value of $C$.
After subtracting this sigmoid fit, we are left with a ``flattened'' profile.
See Fig.~\ref{fig:fitting_proc}  (b).
We fit a double gaussian to the two pre-edge forbidden transitions ($4962~\textrm{eV} < E < 4973~\textrm{eV}$, shown by red dots), and a polynomial to the oscillatory component of the XANES interference features ($E > 4975~\textrm{eV}$, shown by dashed magenta).
These fit components are recombined and an error bar equal to the standard error of the fit combined in quadrature with the error in the crystal reflectivity is added.
See the main text for further discussion.

%--------------------------------------------------------------------------------------%
\bibliography{main} % Produces the bibliography via BibTeX.